\documentclass{PoS}
\usepackage{amsmath}				

\title{SU(2) Lattice Gluon Propagator and Potential Models}

\ShortTitle{SU(2) Lattice Gluon Propagator and Potential Models}

\author{\speaker{Willian Matioli Serenone}%
         \thanks{We acknowledge FAPESP for financial support, grant number 2012/04811-0.}\\
        IFSC, University of São Paulo\\
        CP 369 CEP 13560-970, S\~ao Carlos SP, Brazil\\
        E-mail: \email{willian.matioli@gmail.com}}

\author{Attilio Cucchieri\\
        IFSC, University of São Paulo\\
        CP 369 CEP 13560-970, S\~ao Carlos SP, Brazil\\
        E-mail: \email{attilio@ifsc.usp.br}}

\author{Tereza Mendes\\
        IFSC, University of São Paulo\\
        CP 369 CEP 13560-970, S\~ao Carlos SP, Brazil\\
        E-mail: \email{mendes@ifsc.usp.br}}

\abstract{
We study the bottomonium spectrum using a potential model. Our potential
incorporates lattice results for the gluon propagator, obtained from 
simulations of pure SU(2) gauge theory in Landau gauge. 
The mass of the bottom quark is left as a free parameter.
The resulting spectrum is compared to the case of the Coulomb plus Linear 
(or Cornell) Potential.
}

\FullConference{31st International Symposium on Lattice Field Theory - LATTICE 2013\\
		July 29 - August 3, 2013\\
		Mainz, Germany}

\begin{document}

\section{\label{sec:Introduction}Introduction}

The goal of the present study is to obtain the spectrum of bottomonium states
using a potential-model approach, including nonperturbative input from
lattice simulations. The bottomonium spectrum is a rich one, whose spacings 
between fundamental levels are similar to those of the charmonium. This
motivates the description by a common potential model \cite{Patrignani:2012an}. 
Also, the mass of the bottom quark is large in comparison with the system 
energy (the same happens, to a lesser degree, with charmonium). This 
allows us to make a nonrelativistic approximation to the Bethe-Salpeter 
equation used in the description of bound states of two fermions (see 
e.g.\ Ref.\ \cite{Bijtebier:1995md} and \cite{Bernardini:2003ht}).
More precisely, one makes an instantaneous approximation to the wave function 
(which leads to the Salpeter equation), a nonrelativistic approximation to 
the kinetic term in the Hamiltonian and a local approximation to the potential.
After these approximations, the equation reduces to the Schrödinger equation.

We require the potential to respect some physical characteristics of the 
quark-antiquark interactions in QCD. This is usually accomplished by
a combination of two behaviors. The first one refers to the quark-antiquark 
interaction in the one-gluon-exchange (OGE) approximation (related to
quark-antiquark scattering inside the meson) and is of perturbative nature.
The second one, the property of confinement, may be modeled by a linearly rising 
function, inspired by lattice QCD simulations (see Refs.\ 
\cite{DeRujula:1975ge}, \cite{Bali:2005fu} and \cite[Fig.\ 5]{Donnellan:2010mx}).
If the free vector-boson propagator is used, the obtained potential will be
the ``Coulomb plus linear'' or Cornell potential
\begin{equation}
\label{Cornell}
V(r)\;=\; -\frac{4}{3}\frac{\alpha_s}{r} \,+\, F_0\,r\,.
\end{equation}

Other commonly used potentials are listed in \cite[Section 7]{Lucha:1991vn}. 
Here we substitute the free gluon propagator with the one provided in 
Ref.\ \cite{Cucchieri:2011ig}, obtained from lattice simulations of pure SU(2) gauge 
theory in Landau gauge. This introduces nonperturbative features into
the scattering term of the potential.

In Section \ref{sec:Pot_Model_Review} we briefly describe the procedure
for calculating the potential from the gluon propagator.
In Section \ref{section:Methods} we detail our method for solving the 
Schrödinger equation for an arbitrary potential and therefore obtaining 
the bottomonium spectrum. The results of this method are presented in Section 
\ref{section:Results} and we present our conclusions in Section 
\ref{section:Conclusion}. Preliminary results of our study were reported in
\cite{Serenone:2012yta}.

\section{Brief Review of Potential Models}
\label{sec:Pot_Model_Review}

The use of potential models in the study of heavy quarkonia is based on the 
assumption that the interaction between a heavy quark (namely the charm or the 
bottom quark) and its antiquark may be described by a potential.
This is inspired by the fact that the Coulomb potential, which may be 
obtained as a limiting case from QED, explains with great accuracy 
bound states of nonrelativistic systems such as atoms or the positronium. 

\begin{figure}[htbp]
\begin{minipage}[b]{0.45\textwidth}
    \centering
    \includegraphics[width=0.7\columnwidth]{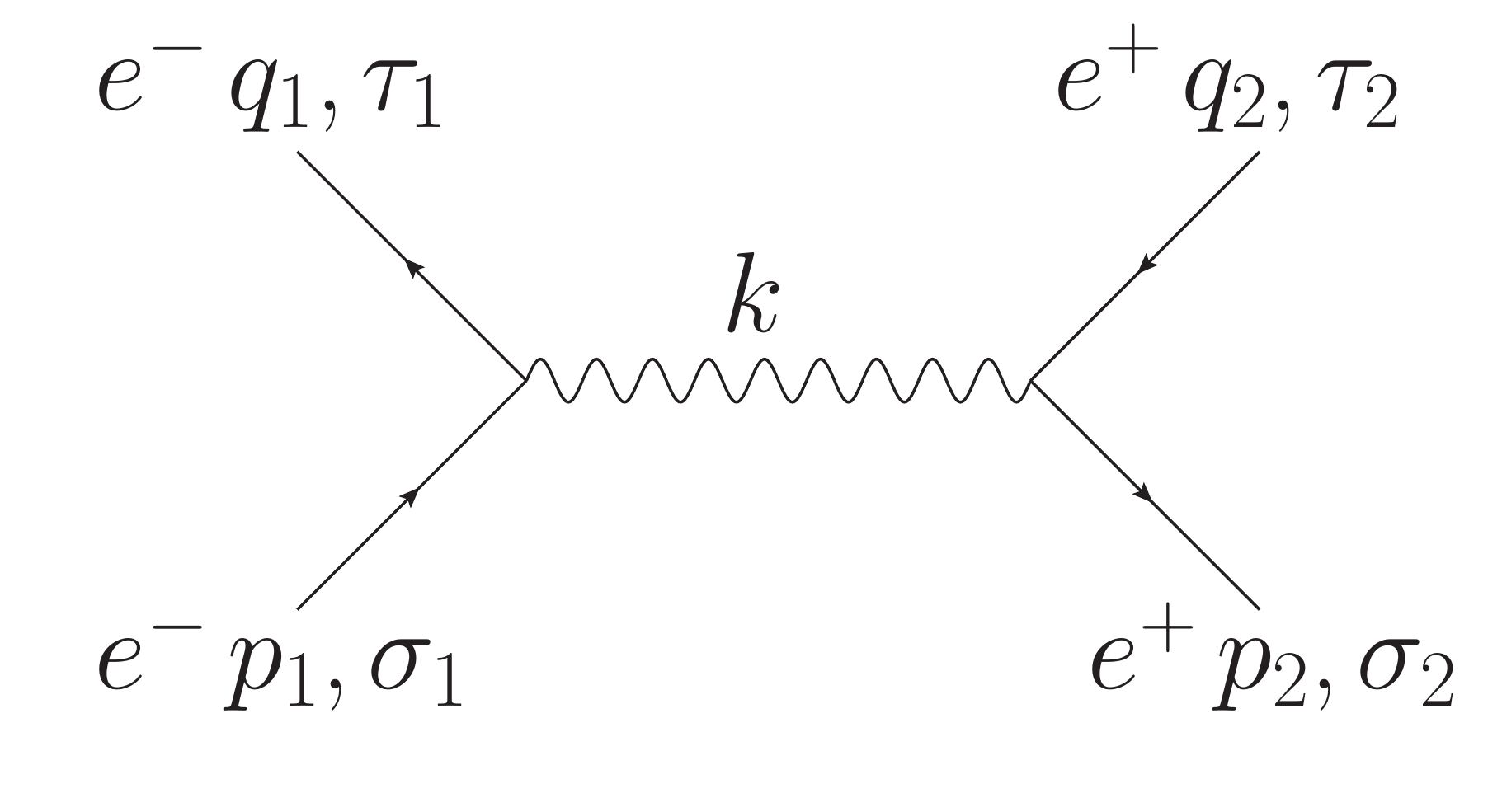}
	\label{fig:Feynman_Graph_1_Electrons}
\end{minipage}\hfill
\begin{minipage}[b]{0.45\textwidth}
\centering
		\includegraphics[width=0.40\columnwidth]{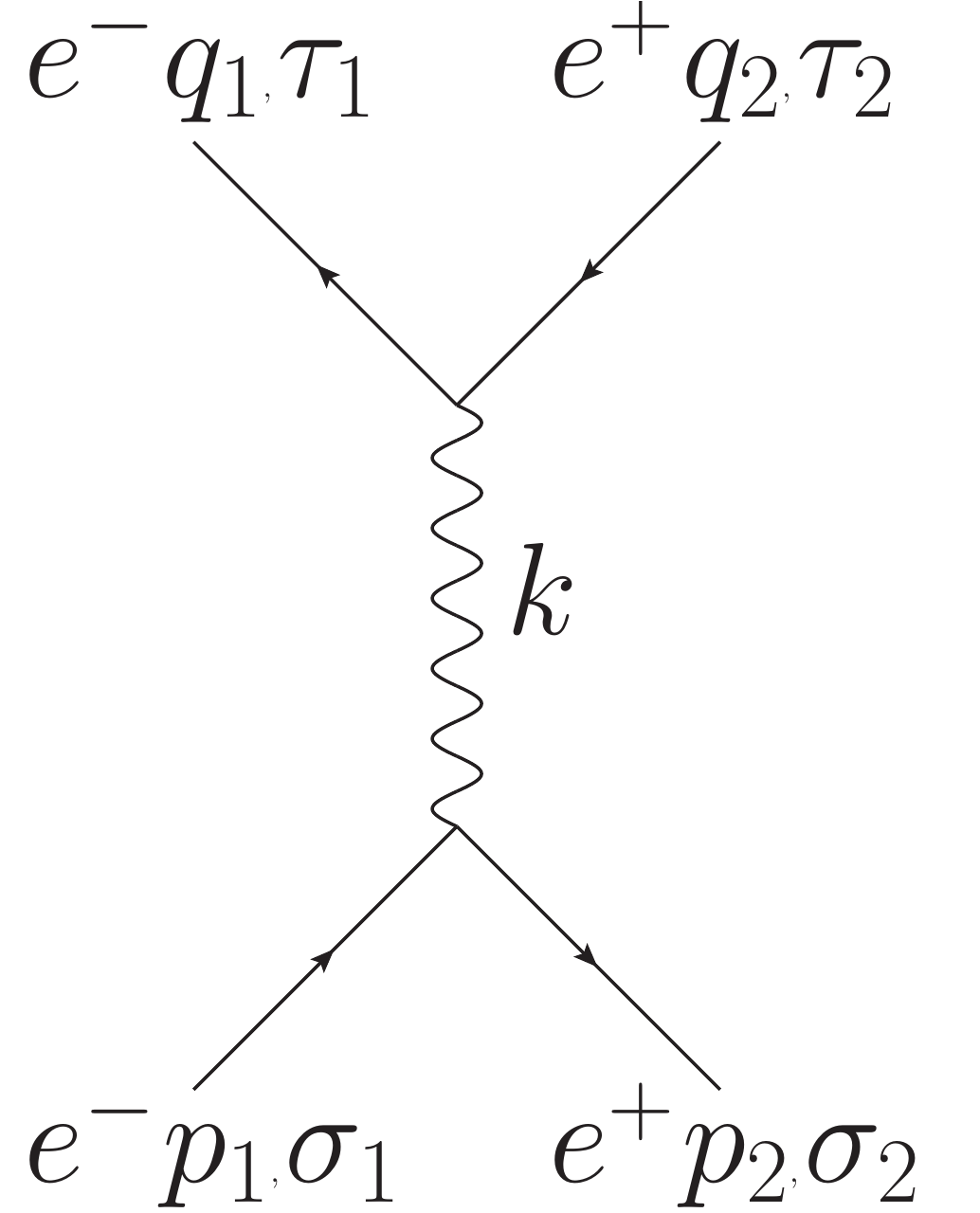}
	\label{fig:Feynman_Graph_2_Electrons}
\end{minipage}
\caption{Feynman diagrams leading to the Coulomb potential in QED.}
\label{fig:Feynman_Graph_Electrons}
\end{figure}

To obtain the Coulomb potential, one starts by considering an {\em elastic} 
$e^-e^+$ scattering process. Applying perturbation theory in the first-order 
Born approximation, we obtain the scattering-matrix element $S_{fi}$
\begin{equation}
 S_{fi} \;\equiv\; \langle f | i \rangle \;=\; \delta_{fi} \,+\, 
i (2 \pi)^4 \, \delta^{(4)}(Q-P)\, T_{fi} \,,
\end{equation}
where $Q$ and $P$ correspond respectively to the final and initial total
momentum and $T_{fi}$ is the scattering amplitude, which can be computed 
through Feynman rules. 
There are two Feynman diagrams contributing to it, which 
are shown in Fig.\ \ref{fig:Feynman_Graph_Electrons}.

These diagrams result in the following scattering amplitude
\begin{align}
 \label{eq:QED_Scattering_Amplitude}
 T_{fi} \;=\; \frac{1}{(2 \pi)^6} \frac{m^2}{
    \sqrt{E_{p_1}E_{p_2}E_{q_1}E_{q_2}}} & \left[
  \,-\,e^2 \,\overline{u}(q_1,\tau_1)\,\gamma^\mu 
  \,u(p_1,\sigma_1)\;P_{\mu \nu}(k)\;\overline{v} 
  (p_2,\sigma_2)\,\gamma^\nu\,v(q_2,\tau_2) \right. \nonumber \\[1mm]
  & \left. \;\,+\,e^2\,\overline{v}(p_2,\sigma_2)\,\gamma^\mu
  \,u(p_1,\sigma_1)\;P_{\mu \nu}(k)\;\overline{u}
  (q_1,\tau_1)\,\gamma^\nu v(q_2,\tau_2)\,\right] \;,
\end{align}
where we follow the notation in \cite{Lucha:1991vn}.

We then make the nonrelativistic approximation, i.e.\ we impose the kinetic 
energy of the system to be much smaller than its rest energy 
($ p \ll m \cong E$).
If the usual photon propagator $P_{\mu \nu}(k) = g_{\mu \nu}/k^2$ is used, 
we see that the contribution of the second term on the r.h.s.\,of 
Eq.\,(\ref{eq:QED_Scattering_Amplitude}), which comes from the annihilation 
diagram, is negligible compared to the first term. 
Thus, performing a spatial Fourier transform, we recover the familiar 
Coulomb potential
\begin{equation}
 V(\textbf{r}) \;=\; - (2 \pi)^3 \int \exp(-i \textbf{k}\cdot\textbf{r})\,T_{fi}(k)\,d^3{r}\;=\;-\frac{1}{(2 \pi)^3} \int \exp(-i \textbf{k}\cdot\textbf{r})\,\frac{e^2}{\textbf{k}^2}\,d^3 r\;=\;- \frac{e^2}{r}\,.
\end{equation}

This potential may be used to model the interaction in the positronium. 
Since we know that the binding energies will be of the order of eV, while the 
electron and positron masses are approximately 0.5 MeV, we can expect that 
the nonrelativistic approximations for the potential will hold. We may use 
this potential in the Schrödinger equation to obtain the 
energy spectrum of the system. Notice that the nonrelativistic 
approximation completely removes spin dependencies of the potential.

In our work we follow the same procedure, replacing the electron-positron
pair by a quark-antiquark pair and the photon by a gluon. For the gluon 
propagator we use an expression obtained from fits of lattice data for
pure SU(2) gauge theory in Landau gauge, given in Ref.\ \cite{Cucchieri:2011ig}. 
The limitation of this
method is that, since we apply perturbation theory to obtain the potential,
$V(r)$ is not expected to be a confining potential, even though the propagator used 
is obtained nonperturbatively. We model confinement, as usual, by adding to
the potential a linearly rising term $F_0 r$.

\section{Method for Obtaining Bottomonium Masses}
\label{section:Methods}

We expect that a typical choice of propagator will result in a 
central potential 
if we use the approximations in Section \ref{sec:Pot_Model_Review}. 
Since our system contains only two particles, the Hamiltonian will be 
essentially the same as the one of the hydrogen atom, written in terms of 
relative coordinates. We use separation of variables to isolate the angular dependence of the wave function (which will be given by the
spherical harmonics) and the variable substitution
$R(r) = f(r)/r$ to obtain the Ordinary Differential Equation (ODE) for $f(r)$
\begin{equation}
\label{eq:Radial_Equation}
\frac{d^2 f}{dr^2}\,+\,2\mu\left[E - V(r) -2m_b - \frac {l\left(l+1\right)}{2 \mu r^2} \right] f\left(r\right) \;=\; 0 \,,
\end{equation}
where $\mu$ is the reduced mass and $m_b$ is the mass of the bottom quark.
Notice the addition of the rest mass of the particles, which will allow us to compare the eigenenergy directly with masses in Ref.\,\cite{Beringer:1900zz}.

Since the potential is arbitrary, it will not usually be possible to find an 
analytic expression for the eigenenergies. 
We therefore use a numerical approach. The algorithm consists in the 
following steps:
\begin{enumerate}
  \item Finding a likely range for the eigenenergies and discretizing this 
interval in $N$ steps separated by $dE$. (We fix the range using the 
experimental values for the lowest and the highest energy states.)
  \item Numerically solving the ODE in Eq.\ (\ref{eq:Radial_Equation}) 
to obtain the function $f(r)$ for each energy. We use the Numerov method 
(see Ref.\ \cite[Chapter 3]{koonin1998computational}).
  \item Estimating the eigenenergy using the boundary conditions. The functions $f(r)$ will generally diverge to $\pm \infty$, since our guess for $E$ in Eq.\,(\ref{eq:Radial_Equation}) is not an eigenenergy. 
If we find that the sign of this divergence is reversed when changing from
$E_{n,i}$ to $E_{n+1,i}$, the $i$th eigenenergy will be estimated by $(E_{n,i}+E_{n+1,i})/2$. The error is taken as $dE/2$.
\end{enumerate}

Note that the only free parameter in the potential is the string force $F_0$,
but we also leave free the mass of the bottom quark since, at present,
it is not well determined. In fact, different approaches give different 
results for $m_b$ (for instance, Ref.\ \cite{Beringer:1900zz} has two values for it). 
To find the best values for these parameters, we adopt a similar strategy 
used in the calculation of the eigenenergies described above: we set a 
range where it is believed the values of the parameters may be and discretize 
it. We then compute the eigenenergies for each proposed set of parameters and select 
the one that best describes the observed spectrum. The criteria for choosing 
a set of parameters with this property is to look for the set that minimizes 
the residual
\begin{equation}
 R(\text{Parameters}) \;=\; \sum_{i} (E_{i}-E_{i,\text{Experimental}})^2\,.
\end{equation}

\section{Results}
\label{section:Results}

Following the discussion in Section \ref{sec:Pot_Model_Review},
the Feynman diagrams for the bottomonium will be similar to the ones in
the $e^- e^+$ scattering shown in Fig.\ \ref{fig:Feynman_Graph_Electrons}.
This similarity means that the scattering matrix will have the same structure. 
The main difference will be extra factors due to the $SU(3)$ symmetry of QCD.
Considering that the gluon propagator is color-diagonal, 
i.e.\ $P^{a b}_{\mu \nu}(k) \propto \delta^{a b}$,
the color factors contributing to the scattering diagram 
and the annihilation diagram will be respectively
\begin{align}
\label{eq:color_factor}
c^\dag_{1,f}\,\frac{\lambda^a}{2}\,c_{1,i}\;c^\dag_{2,i}\,\frac{\lambda^a}{2}\,c_{2,f}\;=\;\frac{4}{3}\\
c^\dag_{2,i}\,\frac{\lambda^a}{2}\,c_{1,i}\;c^\dag_{1,f}\,\frac{\lambda^a}{2}\,c_{2,f}\;=\;0\,,
\end{align}
where the vectors $c_{1,2}$ represent the color states and $\lambda^a$ are the
Gell-Mann matrices.
Notice that here the contribution from the annihilation diagram is exactly zero.

The propagator from Ref.\ \cite{Cucchieri:2011ig} has the form
\begin{equation}
 P_{\mu \nu}^{a b}(k) = \frac{C\,(s^2+k^2)}{t^2+u^2 k^2+k^4}\left(\delta_{\mu \nu} - \frac{k_\mu k_\nu}{k^2}\right) \delta^{a b}\,.
\end{equation}
(Values for the normalization constant $C$ and the parameters $s$, $t$, $u$
are given in \cite{Cucchieri:2011ig}.)

By analogy with the QED case considered before, we keep only the first
term of the above tensor structure.
Also, this
propagator is obtained for Euclidean time and therefore we need to make the 
transformation $\delta_{\mu \nu}\rightarrow g_{\mu \nu}$. (Without this 
transformation the perturbative term would be repulsive.) Furthermore, we approximate 
all energies to the particle mass, which implies $m^2 /\sqrt{E_{p_1}E_{p_2}E_{q_1}E_{q_2}} = 1$. With these approximations, we have
\begin{equation}
T_{fi}(k) =  -\frac{1}{(2 \pi)^6} 
              \left[ \frac{4}{3} g_s^2 \overline{u} (q_1,\tau_1) \gamma^\mu u(p_1,\sigma_1) g_{\mu \nu} \frac{C(s^2+k^2)}{t^2+u^2 k^2+k^4} \overline{v} (p_2,\sigma_2) \gamma^\nu v(q_2,\tau_2) \right]\,.
\end{equation}

We need now to compute the factors coming from the spinors. When we impose 
that the particles be nonrelativistic, we obtain
\begin{align}
\overline{u} (q_1,\tau_1)\, \gamma^\mu \,u(p_1,\sigma_1) &\;=\; \delta^{\mu 0}\, \delta_{\sigma_1 \tau_1}\\
\overline{v} (p_2,\sigma_2)\, \gamma^\nu \,v(q_2,\tau_2) &\;=\; \delta^{\nu 0}\, \delta_{\sigma_2 \tau_2}\,. 
\end{align}
The final scattering amplitude is
\begin{equation}
\label{eq:scattering_amplitude}
T_{fi}(k) \;=\;  \frac{1}{(2 \pi)^6} \left[ \frac{4}{3} \,g_s^2 \,\frac{C(s^2+k^2)}{t^2+u^2 k^2+k^4} \right]\,.
\end{equation}

We proceed to computing the Fourier transform. The angular integral is easily solved by assuming that the point $\textbf{r}$ lies on the z-axis. The radial integration can be easily computed through the residue method. Notice that Eq.\ (\ref{eq:scattering_amplitude}) has four simple poles, one in each quadrant of the complex plane. The angular integration does not add any new pole to it. These poles are distributed in such a way that, once one of them is obtained, it is possible to reproduce all others by complex conjugation and/or multiplication by $-1$. These symmetries allow one to express the four terms coming from the residue calculation around each pole as a single term, dependent only on the pole of the first quadrant. The potential becomes
\begin{align}
V(r) & \;=\;  -\frac{4}{3}\frac{g_s^2}{\pi^2 r}\, \Re\left[\frac{C(s+k_1^2) \exp(i k_1 r)}{4k_1^2-u^4}\right]\,,  \nonumber \\[2mm]
k_1  & \;=\;  i\sqrt{t} \,\exp\left[-\frac{i}{2}\arctan\left(\frac{\sqrt{4t^2-u^4}}{u^4}\right)\right]\,.
\label{eq:Potential_Lattice}
\end{align}

\begin{figure}[htbp]
\begin{minipage}[b]{0.50\textwidth}
	\centering
		\includegraphics[width=.95\columnwidth]{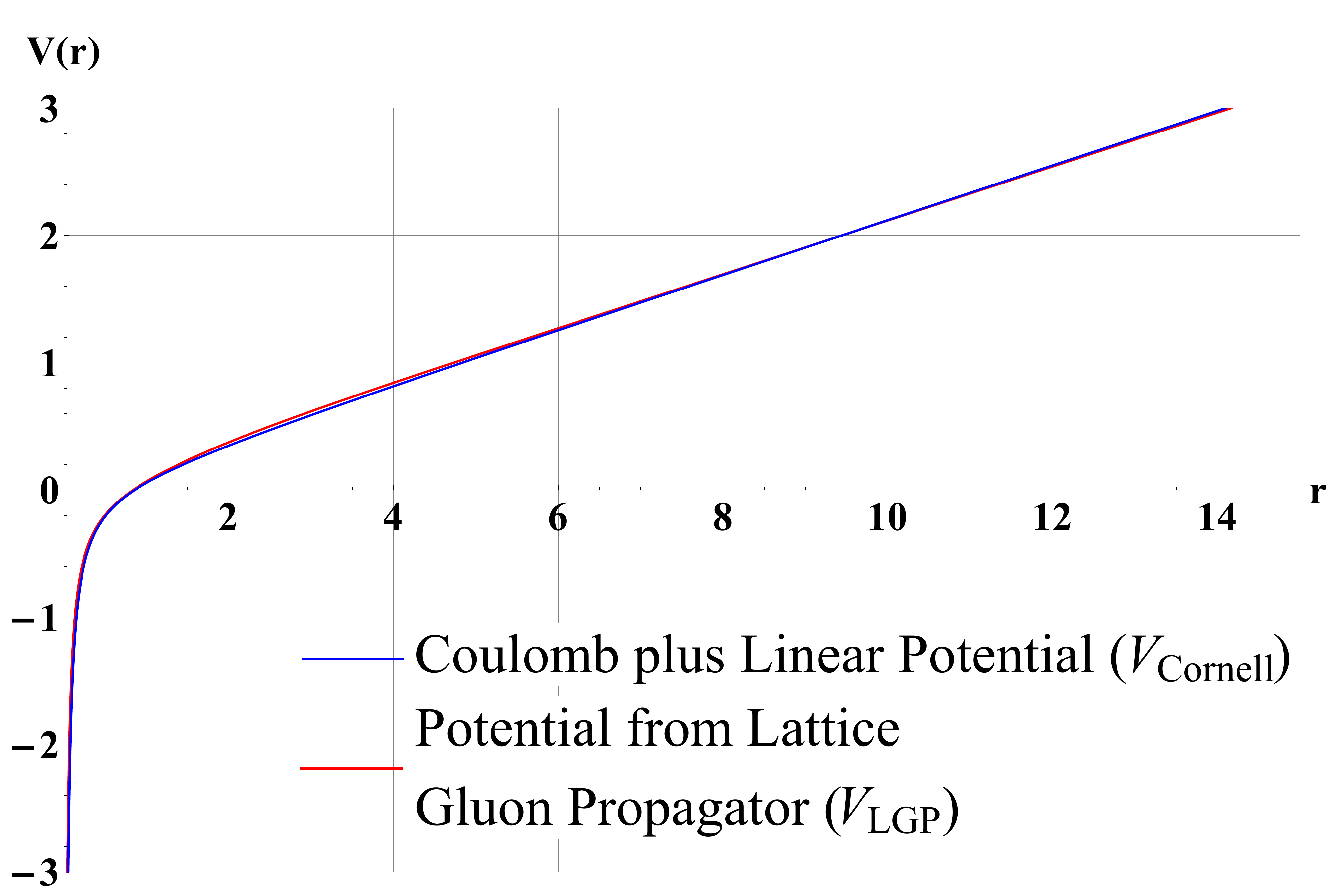}

\end{minipage}\hfill\hspace*{-7mm}
\begin{minipage}[b]{0.55\textwidth}
\centering
		\includegraphics[width=.95\columnwidth]{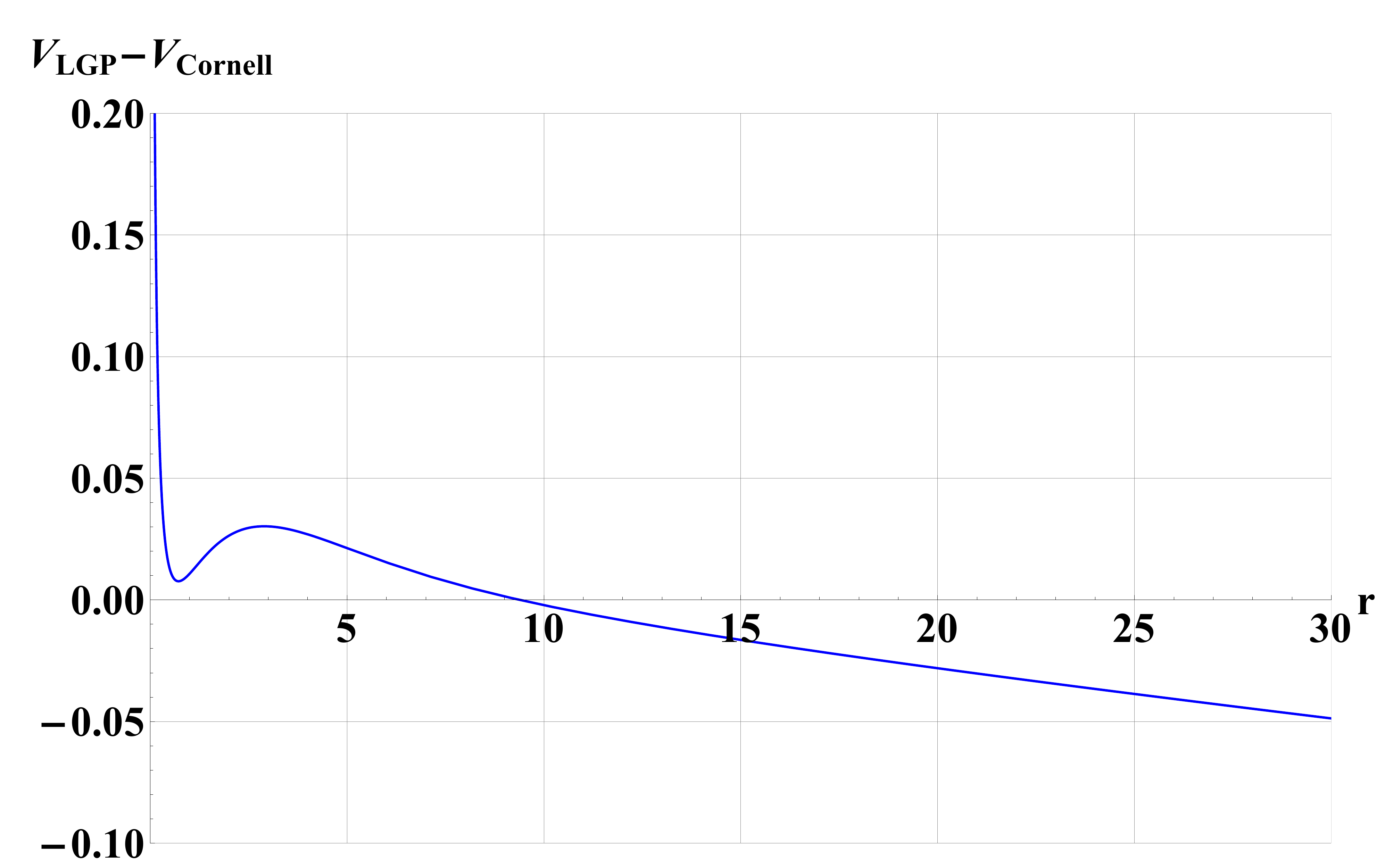}

\end{minipage}
\caption{Plot of the potential obtained using the lattice gluon propagator 
and comparison with the Cornell potential (left panel). The computed string 
tension is only slightly different for the two cases. 
We also plot the difference between these two potentials (right panel). In
both cases the potentials are given in GeV and the distances $r$ in GeV$^{-1}$.
}
\label{fig:Potential}
\end{figure}

We implement the algorithm described in Section \ref{section:Methods} for quark masses from 4.1 GeV through 4.8 GeV (which includes
both masses listed in Ref.\ \cite{Beringer:1900zz}). For the string-tension parameter $F_0$, we search from 0.1 GeV$^2$ through 0.3 GeV$^2$. We apply the same algorithm using the Coulomb plus linear potential for comparison. The results can be found in Table \ref{table:Results}.
\begin{center}
\begin{table}[h]
\resizebox{15cm}{!}{
\begin{tabular}{||c|c||c|c||c|c||}
\hline
\multicolumn{ 2}{||c||}{$m_b$($\overline{\text{MS}}$) = $4.18(3)$ GeV} & \multicolumn{ 1}{c}{Potential from}     & $F_0 = 0.2118(1)$ GeV$^2$      & \multicolumn{ 1}{c}{Coulomb plus}    & $F_0= 0.2136(1)$ GeV$^2$ \\

\multicolumn{ 2}{||c||}{$m_b$(1S)$ = 4.66(3)$ GeV}                   & \multicolumn{ 1}{c}{Lattice Propagator} & $m_b = 4.5977(1)$ GeV         & \multicolumn{ 1}{c}{Linear Potential} & $m_b = 4.6090(1)$ GeV \\

\multicolumn{ 2}{||c||}{See \cite{Beringer:1900zz}}                                      & \multicolumn{ 1}{c}{}                   & $R = 0.0436$                          & \multicolumn{ 1}{c}{}               & $R  = 0.0475$ \\
\hline
{\bf Particle} & \multicolumn{ 1}{c||}{{\bf Experimental}} & {\bf Calculated Mass} & \multicolumn{ 1}{c||}{{\bf Deviation from}}         & {\bf Calculated Mass} & \multicolumn{ 1}{c||}{{\bf Deviation from}} \\

{\bf State}   & \multicolumn{ 1}{c||}{{\bf Mass(GeV)}}   & {\bf (${\bf \pm 3 \times 10^{-4}}$ GeV)}    & \multicolumn{ 1}{c||}{{\bf Experiment (GeV)}} & {\bf (${\bf \pm 3 \times 10^{-4}}$ GeV)}    & \multicolumn{ 1}{c||}{{\bf  Experiment (GeV)}} \\
\hline
       1S* & 9.42565(153) &     9.5763 &     0.1507 &     9.5793 &     0.1528 \\
\hline
        2S & 10.02326(31) &    10.0071 &     0.0162 &    10.0029 &     0.0204 \\
\hline
        3S & 10.3552(5)  &    10.3317 &     0.0235 &    10.3293 &     0.0259 \\
\hline
        4S & 10.5794(12) &    10.6107 &     0.0313 &    10.6119 &     0.0325 \\
\hline
        5S & 10.876(11) &    10.8633 &     0.0127 &    10.8675 &     0.0085 \\
\hline
        6S & 11.019(8) &    11.0973 &     0.0783 &    11.1045 &     0.0855 \\
\hline
       1P* & 9.89076(82) &     9.8595 &     0.0313 &     9.8565 &     0.0343 \\
\hline
       2P* & 10.25410(94) &    10.2033 &     0.0508 &    10.2009 &     0.0532 \\
\hline
        3P & 10.530(14) &    10.4943 &     0.0357 &    10.4949 &     0.0351 \\
\hline
        1D & 10.1637(14) &    10.0743 &     0.0894 &    10.0683 &     0.0954 \\
\hline
\end{tabular}
}
\caption{Comparison between the results obtained for the potential extracted 
using the lattice gluon propagator and the usual Coulomb plus Linear potential.
Notice that the states marked with ``*'' are actually
an average of states with different spin but same orbital angular momentum.}
\label{table:Results}
\end{table}
\end{center}
We remark that our obtained values for the quark mass agree much better 
with the $m_b$(1S) value from Ref.\ \cite{Beringer:1900zz}. The small difference between 
the results for our potential and for the Coulomb plus linear potential can traced to 
the fact that the two potentials are nearly identical,
as show in Fig.\ \ref{fig:Potential}.

\section{Conclusion}
\label{section:Conclusion}

We compute the bottomonium spectrum using a potential obtained 
from lattice simulations of the gluon propagator and compare to 
results using the Coulomb plus linear potencial. 
We find very similar behavior, with slightly better agreement 
with experimental data in the former case. This might indicate that, although 
the perturbative treatment removes most of the nonperturbative features of 
the scattering contribution to the potential, a small part of them survives 
this treatment.

We note that the propagator we used was computed for pure 
SU(2) gauge theory, instead of (pure) QCD. (Note also that this propagator 
is determined up to a global constant $C$, fixed by normalization.) 
The good agreement of our results with the experimental spectrum suggests
that it is enough to take into account the SU(3) color structure by
including the usual color factor in the calculation 
(see Eq.\ (\ref{eq:color_factor})).

In our study we do not include effects of spin-spin
interaction or spin-orbit interactions, which can
lead to splitting of some energy levels.
We have as well additional errors due to our 
nonrelativistic approach. The main limitation, however, is clearly the use
of the OGE approximation in the scattering calculation,
and the need to include the confining term by hand. We nevertheless
believe that the method is useful, especially if an application
beyond the OGE approximation can be made.
(We note that a different point of view is presented in a similar study
reported in Ref.\,\cite{Gonzalez:2011zc}.)

\bibliographystyle{JHEP} 
\bibliography{Proceedings_Lattice_2013}

\end{document}